\documentclass[aps,pra,twocolumn,groupaddress,floatfix]{revtex4-1}

\usepackage[latin1]{inputenc}
\usepackage[english]{babel}
\usepackage{setspace, graphicx, amssymb}
\usepackage{amsmath}
\usepackage{amscd}
\usepackage{rotating}
\usepackage{color}
\usepackage{epstopdf}
\usepackage{subfigure}
\usepackage{natbib}

\newcommand{\vect}[1]{\textbf{#1}}

\begin{document}

\title{Two mode coupling in a single ion oscillator via parametric resonance }
\author{Dylan J Gorman\textsuperscript{1}}
\author{Philipp Schindler\textsuperscript{1}}
\author{Sankaranarayanan Selvarajan\textsuperscript{1,2}}
\author{Nikos Daniilidis\textsuperscript{1}}
\author{Hartmut~H\"affner\textsuperscript{1}}
\affiliation{\vspace{.1cm}
\mbox{$^1$Dept. of Physics, University of California, Berkeley, CA 94720, USA}\\
\mbox{$^2$Institut f\"ur Experimentalphysik, Universit\"at
Innsbruck, A-6020 Innsbruck, Austria}
\vspace{5pt}}
\begin{abstract}
Atomic ions, confined in radio-frequency Paul ion traps, are a
promising candidate to host a future quantum information
processor.
%
% In such devices, performing many-body quantum gates requires state
% selective manipulation of the ion's motion, usually performed with
% laser light driving an optical transition. Therefore, providing
% optical access is a major design consideration which becomes
% increasingly difficult for miniaturized trap designs enabling scalable
% quantum computing.
%
In this letter, we demonstrate a method to couple two motional modes of a
single trapped ion, where the coupling mechanism is based on applying
electric fields rather than coupling the ion's motion to a
light field. This reduces the design constraints on the experimental
apparatus considerably. As an application of this mechanism, we cool a
motional mode close to its ground state without accessing it optically.
As a next step, we apply this technique to measure the mode's
heating rate, a crucial parameter determining the trap quality.
In principle, this method can be used to realize a two-mode quantum
parametric amplifier.
\end{abstract}

\maketitle

%%%%%%%%%%%%%%%%%%%%%%%%%%%%%%%%%%%%%%%%%%%%%%%%%%%%%%%%%%%%%%%%%%%%%%%%%%%%%
%%%%%%%%%%%%%%%%%%%%%%%%%%%%%% INTRO %%%%%%%%%%%%%%%%%%%%%%%%%%%%%%%%%%%%%%%%
%%%%%%%%%%%%%%%%%%%%%%%%%%%%%%%%%%%%%%%%%%%%%%%%%%%%%%%%%%%%%%%%%%%%%%%%%%%%%

\section{Introduction}

The past several decades have yielded tremendous progress in
controlling the quantum states of single trapped ions
\cite{Blatt2008,Haeffner2008,leibfried2003,Monz2011}. A crucial aspect of
experimental work in trapped ion physics is the coupling of the
electronic spin to its quantum mechanical motion utilizing transitions
at optical frequencies, enabling laser cooling and many body quantum
gates. However, the coupling strength of such spin-motion transitions
depends crucially on the projection of the laser wavevector onto the
mode of interest ~\cite{leibfried2003}. Thus, motional modes with
small projection onto the laser wavevector remain difficult to control and
interrogate, constraining the design of the ion trap as well as the
entire experimental apparatus.

% The problem can be partially overcome by rotating the normal modes such that the laser has substantial projection onto each mode, allowing direct access to all motional modes at once. However, this imposes additional constraints to the trap design and thus it is advantageous if it would be possible to access motional modes with a different mechanism that is not constrained by laser beam geometry. Additionally, some experiments necessitate a particular orientation of the trap modes which makes certain modes inaccessible. Among those experiments are those aimed at constructing hybrid quantum devices out of trapped ions and solid-state circuits. 

In the following, we describe a parametric coupling scheme allowing
experimental access to any vibrational mode of a single ion without direct
optical interaction. %This work is similar to techniques already
%demonstrated in Penning traps\cite{Wineland1975,cornell1990}.
For this, an oscillating potential is applied to trap electrodes such that
the associated field features a spatial variation enabling the
coupling. We will consider this first in the case of a general two
dimensional harmonic oscillator with the potential
$\frac{m}{2}\left(\omega_x^2 x^2 + \omega_y^2 y^2\right)$. The
introduction of a perturbing potential with a term proportional to
$xy$ will make the instantaneous eigenmodes of the system no longer align
purely in the $x$ and $y$ directions, creating a coupling between
the modes.
%In order for population to completely
%swap, we expect energy to be conserved. This implies that the
%perturbing potential should be modulated at the difference frequency
%$\omega_p = \omega_x - \omega_y$ (WHY???). 
If the two modes have different frequencies, energy conservation considerations
imply that single quanta cannot be exchanged between the two modes if the
coupling potential is static in time. This exchange becomes possible by
modulating the perturbing potential. The energy difference between the two modes is then made up for by either
absorbing a quantum from, or emitting a quantum into, the driving
field. This system features dynamics where the motional energy oscillates between
the the mode aligned with the $x$ axis (the X mode) and the one aligned with the
$y$ axis (the Y mode).

In Paul traps, the ion is effectively confined in a three dimensional
oscillator potential, which arises by applying radio-frequency
(rf) and static voltages to a number of trap electrodes. The trapped
ion's motion can be decomposed into three normal modes. In order to
couple two of the normal modes, we add a time-varying voltage to one or more nearby
electrodes, which oscillates at the difference between the modes' frequencies.

% The trapping potential in Paul traps is generated by two different fields where the first part is a pseudopotential that arises from placing a radio-frequency~(rf) driving potential with a frequency in the several tens of MHz on a set of electrodes. This pseudopotential  is given by the squared electric field amplitude at the trapping position, and gives rise to harmonic trapping when the rf creates a voltage with a quadrupole spatial configuration. The second contribution to the trapping potential is given by the total electrostatic potential at the trapping position, where the interaction energy is given by the product of the ion's charge with the electric voltage. The trapped ion's motion can be decomposed into three normal modes. In the following we add an oscillating voltage to one or more nearby electrodes, which is able to couple two of the normal modes if it oscillates at the difference frequency between the modes.

A similar method of mode-mode coupling was proposed and demonstrated
for charged particles in Penning traps~\cite{Wineland1975,Brown1986,cornell1990}
 %~\cite{wineland1975,cornell1990,Gabrielse??},
where the axial and cyclotron modes of a single charged particle are
coupled by an additional electric field. We apply these methods to
microfabricated surface traps, which are a promising candidate for a
scalable quantum information processor~\cite{Kielpinski2002} but have
inherently limited optical access. We demonstrate that this technique
can be used to cool trap modes lacking substantial overlap with the
wavevector close to their ground state and directly apply this
technique to perform a heating rate measurement without directly
accessing the mode of interest. Thus, this scheme allows an
experimenter to manipulate all motional modes of a single ion,
even when optical access is restricted.

This paper is organized as follows. In Sec. II we discuss the theory of this interaction. Sec. III covers the details of our experiment, and characterizes the coupling process in the frequency domain. Sec. IV includes our experimental results, where we show several applications of this technique. Finally, Sec. V summarizes our results and suggests extensions to this work.

%%%%%%%%%%%%%%%%%%%%%%%%%%%%%%%%%%%%%%%%%%%%%%%%%%%%%%%%%%%%%%%%%%%%%%
%%%%%%%%%%%%%%%%%%%%%%% THEORY %%%%%%%%%%%%%%%%%%%%%%%%%%%%%%%%%%%%%%%
%%%%%%%%%%%%%%%%%%%%%%%%%%%%%%%%%%%%%%%%%%%%%%%%%%%%%%%%%%%%%%%%%%%%%%

\section{The coupling mechanism}

We consider an ion, with charge $q$ and mass $m$, confined in a linear
surface electrode rf Paul trap. In such a trap, an oscillating voltage
with frequency in the range of $2\pi \times 30 \rm{\; MHz}$ will be
applied to two electrodes (labelled ``RF" in Fig. ~\ref{fig:trap})
providing two-dimensional confinement along the X and Y axes. In
general, the motion of an ion in such a trap is a solution to the
Matthieu equation \cite{Paul1990}, but usually the dynamic trapping
effects can be neglected in favor of the ``pseudopotential
approximation" \cite{Dehmelt1967}. In this approximation, the
time-dependent Hamiltonian induced by the trapping rf is replaced with
a time-independent harmonic oscillator with motional frequencies
$\omega_x$ and $\omega_y$. In our case the rf pseudopotential
generates no confinement in a third direction, $z$. The confinement in
the $z$ direction is the result of a dc potential which is
quadratically varying in space, giving the ion a third motional
frequency $\omega_z$. The total Hamiltonian is then treated as the
pseudopotential from the trapping rf field, plus the total
electrostatic potential. In the following, we will consider the
trapping potential static, and the only time varying field is the one
enabling the mode-mode coupling.
% As a rough guideline the motional frequencies tend to be about an
% order of magnitude smaller than the trapping rf frequency. It is
% potentially confusing that both the trapping mechanism and the
% parametric coupling mechanism arise from rf fields created on the
% trap. However, for the rest of this paper we shall always work in
% the pseudopotential approximation, where the trapping Hamiltonian is
% time independent, and the rf we will consider is the one which
% generates parametric coupling.
 
Working in the aforementioned approximation, we consider a harmonically confined ion with motional frequencies $\omega_i$, $i \in \{x, y, z \}$. The interaction energy created when a voltage $V$ is applied to a nearby coupling electrode is $q U(\vect{r}) V$. $U(\vect{r})$ is the potential per volt applied to the electrode, describing the spatial profile of the potential at the trapping position due to the coupling electrode. To enable the mode-mode coupling, we modulate the voltage on a set of electrodes whose spatial profile mixes two of the ion's normal vibrational modes.

Then the Hamiltonian governing the motion is $H = H_{\rm{osc}} + q U(\vect{r}) V$, with $H_{\rm{osc}} = \hbar\sum_i \omega_i a_i ^\dagger a_i$ being the harmonic oscillator Hamiltonian and $a_i$ ($a^\dagger _i$) is the annihilation (creation) operator for mode $i$. To achieve mode-mode coupling, we apply an oscillating radio-frequency (rf) voltage on a judiciously chosen set of coupling electrodes such that $V = V_0 \cos \left( \omega_p t \right)$. If $\omega_p = \omega_i - \omega_j$, the difference frequency between modes $i$ and $j$, a parametric coupling emerges in the Hamiltonian.

To see the coupling explicitly, $U(\vect{r})$ is expanded to second order as $U(\vect{r}) = U(0) + \sum_{i}(r_i/D_{1, i}) + (1/2)\sum_{i,j} (\pm) (r_i r_j )/D^2_{ij}$. The linear terms create an electric field at the ion position and present a driving force on the ion which introduces a driven motion analogous to micro-motion. As we will show later, this additional term does not alter the coupling dynamics and thus can be neglected if the set of coupling electrodes are chosen such that the coupling dominates over this driving force.

The terms proportional to $r_i^2$ modify the motional frequencies of the ion. If the modulation frequency is near the resonance condition $\omega_p \approx 2 \omega_i$, these terms effect a parametric amplification of the energy in the $\omega_i$ mode~\cite{Meekhof1996}. However, if $\omega_p$ is far from this condition (as will be the case in our experiments), the modulation of the trap frequency contributes only an overall phase to the ion's spatial wavefunction. Finally, the cross terms proportional to $r_i r_j$ are responsible for the parametric coupling with the drive frequency chosen appropriately.

In the interaction picture, the Hamiltonian becomes 
% \begin{align}
% H &= q \cos(\omega_p t)\left( V_0 \sum_{i,j} \left( \frac{r_i r_j}{2 D^2_{i j}} \right) + \vect{r} \cdot \vect{E} \right) \; .
% \label{eqn:full_H}
% \end{align}
% Here we have defined $\vect{E}$ as $E_i = V_0/D_{1,i}$, representing the electric field at the ion position due to the drive. We will put this term aside for the moment and focus on the term generating the parametric coupling $H_I$ which is in the interaction picture:
\begin{widetext}
\begin{align}
H_I &=  q  V_0  \cos(\omega_p t)\sum_{i,j} \left( \frac{r_i r_j}{2 D^2_{i j}} \right) \nonumber \\
&= q V_0\hbar   \cos(\omega_p t) \sum_{ij} \frac{ e^{ i (\omega_i + \omega_j)t} a^\dagger_i a^\dagger_j + e^{  i (\omega_i - \omega_j)t}a^\dagger_i a_j  + \rm{H.c.} } {4 m \sqrt{\omega_i \omega_j}D^2_{ij}} \nonumber \\
&= \hbar \cos(\omega_p t) \sum_{ij}g_{ij}\left(e^{ i (\omega_i + \omega_j)t} a^\dagger_i a^\dagger_j + e^{  i (\omega_i - \omega_j)t}a^\dagger_i a_j  + \rm{H.c.} \right)
\label{eqn:full_td_int}
\end{align}
\end{widetext}
where $\rm{H.c.}$ indicates Hermitian conjugation. In the last line we have absorbed all the constants except $\hbar$ into the coupling frequency $g_{ij}$. In general, we expect the rotating wave approximation (RWA) to be valid whenever $g_{ij} \ll \omega_{i,j}$.
% In the experiments described here, the motional frequencies are around $2 \pi \times 1 \rm{\; MHz}$ or higher, and the coupling frequencies are less than $10 \rm{\; kHz}$, so we expect the rotating wave picture to be quite accurate in this regime.

If $\omega_p = \omega_i - \omega_j$ and applying the RWA, all the terms in the sum of Eq.~\ref{eqn:full_td_int} vanish except the one involving coupling oscillators $i$ and $j$ leading to
\begin{align}
H_I &\approx \hbar g_{ij} (a_i a^\dagger_j + a^\dagger_i a_j) \; .
\label{eqn:hcoupling_rwa}
\end{align}
This is precisely the interaction we have sought to create: it will swap the quantum states between oscillator modes $i$ and $j$ at a frequency $g_{ij}$. By applying the parametric drive for a specific duration, we can controllably induce state swapping between any two modes of the single ion oscillator.

When the parametric drive is operated on resonance, that is, $\omega_p = \omega_i - \omega_j$, the interaction picture Hamiltonian is diagonal in the basis of two modified normal modes given by $\frac{1}{\sqrt{2}}(a_i \pm a_j)$. The modes are split in frequency by $2g_{ij}$.
If the drive is detuned by $\Delta$ from the parametric resonance, the form of the interaction picture Hamiltonian changes. To treat this problem, it is easiest to transform to a particular interaction picture in which:
\begin{align}
H_I &\approx \hbar \frac{\Delta}{2}(a_i^\dagger a_i - a_j^\dagger a_j) + \hbar g_{ij} (a_i a^\dagger_j + a^\dagger_i a_j) \; .
\label{eqn:hcoupling_delta}
\end{align}
The eigenvalue splitting of this Hamiltonian is given by $2\sqrt{g^2_{ij} + 4\Delta^2}$. Thus, optical spectroscopy of the ion motion will show the bare resonance at $\omega_i$ split into two lines as the parametric drive is operated near resonance. Varying both laser frequency and the parametric drive detuning will show a familiar avoided crossing behavior, providing a witness of the parametric interaction.

In addition to driving the system at the difference frequency between the wo modes,
one can also drive the system at the sum frequency. In that case, Eq.~\ref{eqn:full_td_int}  becomes
\begin{align}
H_I &\approx \hbar g_{ij} (a^\dagger_i a^\dagger_j + a_i a_j)  \; .
\label{eqn:hamplifier_rwa}
\end{align}
in the RWA. This means that the system can be operated as a parametric amplifier if it is driven by the sum frequency of the two modes.

%It should be noted that the above result applies to the special case when a single electrode is driven at frequency $\omega_p$. However for technical reasons it might be necessary to drive multiple electrodes simultaneously. The same solution applies in the general case when  a set of electrodes, indexed by $k$ are all driven in phase with voltage $\alpha_k V$. For spatial profile of the potential due to electrode $k$ of $U_k(\vect{r})$ the interaction energy between the ion and the driving field is given by $V \sum_k \alpha_k U_k(\vect{r})$. Then by making the identification $U(\vect{r}) = \sum_k \alpha_k U_k(\vect{r})$, all of the above results hold without modification.

%%%%%%%%%%%%%%%%%%%%%%%%%%%%%%%%%%%%%%%%%%%%%%%%%%%%%%%%%%%%%%%%%%%
%%%%%%%%%%%%%%%%%%%%%%%%%%%%%%%%%%%%%%%%%%%%%%%%%%%%%%%%%%%%%%%%%%%
%%%%%%%%%%%%%%%%%% EXPERIMENTAL IMPLEMENTATION %%%%%%%%%%%%%%%%%%%%
%%%%%%%%%%%%%%%%%%%%%%%%%%%%%%%%%%%%%%%%%%%%%%%%%%%%%%%%%%%%%%%%%%%
%%%%%%%%%%%%%%%%%%%%%%%%%%%%%%%%%%%%%%%%%%%%%%%%%%%%%%%%%%%%%%%%%%%

\section{Experimental implementation}

\begin{figure}[htbp] 
	\begin{center}
	\includegraphics[width=1\columnwidth]{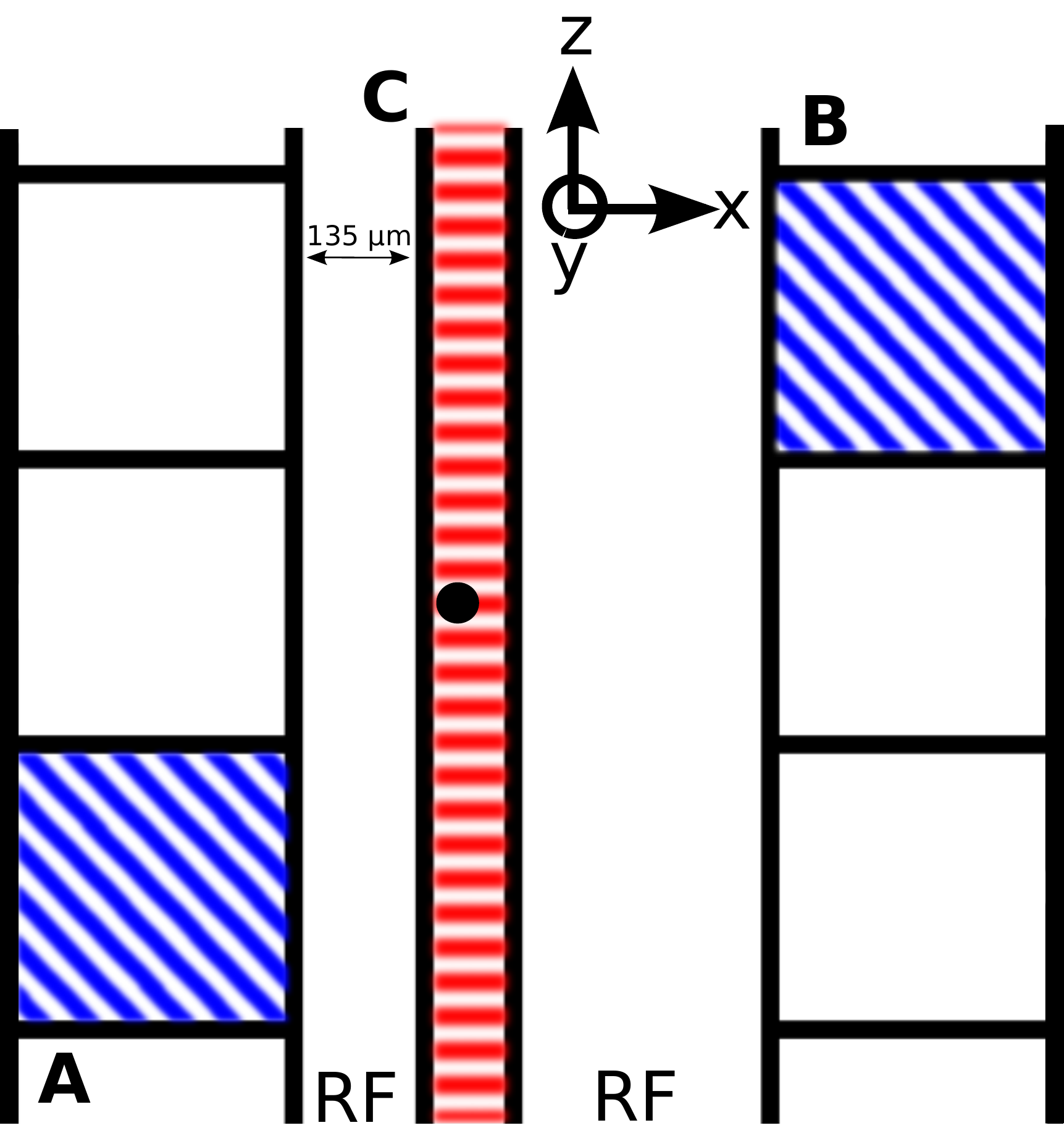}
        \caption{Illustration of the surface trap where the ion's position is
          represented by the black dot. When the experiment is operated in the
          $xz$ coupling configuration, the rf parametric drive is applied to
          the electrodes labeled A and B (blue diagonal shading). When
          operated in the $xy$ coupling configuration, the drive is applied to
          the electrode labeled C (red horizontal shading). }
	\label{fig:trap}
	\end{center}
\end{figure}

In our experiment, a single $^{40}$Ca$^+$ ion is trapped about $100 \rm{\; \mu m}$ above the surface of a micro fabricated surface electrode rf Paul trap, where sideband-cooling and analysis of the motional state is performed on the metastable $4^2S_{1/2} \leftrightarrow 3{^2}D_{5/2}$ transition. The ion has three motional modes, with axes nearly parallel to the Cartesian axes defined in Fig.~\ref{fig:trap}. The motional frequencies along these three axes are about $( \omega_x, \omega_y, \omega_z) \approx 2\pi \times (2.6 \rm{\; MHz}, 2.9 \rm{\; MHz}, 1.0 \rm{\; MHz})$. Sideband cooling and state manipulation is accomplished with a laser in the x-z plane, with about $45^{\circ}$ projection onto both the x and z axes. The projection of the wavevector onto the y axis is $9^{\circ}$, making the Y motional mode difficult to analyze directly. Owing to the small coupling strength of the  laser onto the Y mode, sideband cooling close to the ground state can only be performed on the X and Z modes.

Depending on which modes ought to be coupled, the field needs to be
applied to a set of electrodes maximizing the coupling term while
keeping the linear terms sufficiently small. Throughout this letter we
use two configurations that couple either the X and Y modes or the X
and Z modes. For the first (xy) configuration we simply apply the
coupling field to the electrode marked C in Fig~\ref{fig:trap}. In the
case of the xz configuration, driving a single electrode is not
sufficient as it would result in excessive driving force on the ion (details in Sec. III(a)). 
Therefore, we aim
to apply voltages with ratio of 1:4 to electrodes A and B, which
constitutes the optimal configuration when being constrained to
driving two electrodes in phase.

Each experiment begins with with Doppler cooling on the $4^2S_{1/2} \leftrightarrow 4{^2}P_{1/2}$ transition and optical pumping into the $m_s = -1/2$ state, followed by a fixed length coherent excitation pulse and electron shelving state readout \cite{Wineland1998,Dehmelt1975}. The Doppler cooling stage prepares the Z mode to a mean occupation of $\approx$ 20 vibrational quanta, and the X mode to $\approx$ 6 quanta. The spectroscopy is carried out on the  $|L, m_J\rangle = |S, -1/2\rangle \rightarrow |D, -1/2\rangle$  transition. The state of the motional mode $i$ is probed by evaluating the strength of the sidebands of this transition detuned by $\pm \omega_i$ \cite{Turchette2000}.

The parametric interaction is studied in two ways. It is first
characterized by operating it in \emph{continuous-wave} (CW) mode. In
this mode, the drive is active throughout the experiments and the
spectroscopic signatures of coupling are observed. It is also operated
in \emph{pulsed mode} where the coupling field is switched on for a
fixed time after the initial state preparation giving access to the
time dynamics of the coupling process.

%% In order to cool a mode without being able to perform sideband cooling
%% one first cools another, accessible mode and performs a coupling
%% operation that interchanges the motional states of the two modes. If
%% it is possible to directly spectroscopically analyze the state of the
%% weak mode, this is sufficient to measure a heating rate on this
%% mode. Otherwise, a second coupling operation has to be performed,
%% allowing for spectroscopic analysis on the strong mode.

% In order to measure a heating rate without directly accessing the
% motional mode, a second coupling pulse is applied after a wating time
% $\tau$. The number and placement of the parametric coupling pulses
% varies depending on the application, as discussed in Section IV.

%The rest of this letter will be organized into several parts. First,
%we characterize the coupling with the parametric drive operated in CW
%mode. We spectroscopically observe normal mode splitting when the
%parametric drive is operated near resonance in order determine the
%coupling strength $g_{ij}$. In Sec. IV we then demonstrate the pulsed
%operation of the parametric drive where we ground state cool a single
%mode and then scan the coupling time to directly witness the
%population swapping between a Doppler cooled mode and a ground state
%cooled mode. Finally, we consider two applications of this technique:
%simultaneous ground state cooling of two modes, and measuring the
%heating rate of a mode without interaction with the light field.

\subsection{Characterization of the parametric drive interaction}

The experimentally simplest way to investigate the parametric interaction is to first operate it in CW mode near the parametric resonance condition. Then, laser spectroscopy near one of the secular sidebands (indexed by $i$ or $j$) will show two Lorentzian lineshapes split in frequency space by $2\sqrt{g^2_{ij} + 4\Delta^2}$, where $\Delta$ is the detuning of the parametric drive from the resonance condition.

\begin{figure}[htbp] 
	\begin{center}
	\includegraphics[width=1\columnwidth]{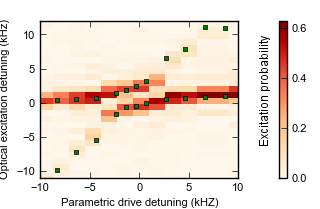}

        \caption{Measured energy spectrum of the X radial sideband
          illustrating the avoided crossing as a function of the
          detuning of the parametric drive. The green rectangles
          represent the mean values of Lorentzian fits to determine
          the frequency splitting. }
	\label{fig:avoided_crossing}
	\end{center}
\end{figure}

By measuring the spectrum around the sideband transition for several
drive frequencies $\omega_p$, the precise resonance frequency and the
coupling strength can be determined from the avoided crossing as shown
in Fig.~[\ref{fig:avoided_crossing}]. At parametric resonance, the
line splitting features a minimum, and the coupling strength equals half
the on-resonance splitting. 
% In principle, the maximum coupling
% strength is limited by validity of the rotating wave approximation,
% wherein we require the coupling frequency to be small when compared to
% the secular frequencies. 
In our setup, coupling strength is limited by the maximum voltage that
can be applied to the coupling electrodes which are heavily filtered
by in-vacuum low-pass filters to suppress heating from technical noise
sources ~\footnote{AVX X7R 47nF; part no. W3H15C4738AT1F showing a
  measured cut-off frequency of 300~kHz}. Nevertheless, we have been
able to achieve coupling frequencies approaching $2 \pi \times 10$~kHz
for both xz and xy coupling configurations.

%\begin{figure}[htbp] 
%	\begin{center}
%	\includegraphics[width=3in]{splitting_detuning_fit.pdf}
%	\caption{ }
%	\label{fig:splitting_detuning}
%	\end{center}
%\end{figure}

This coupling frequency should be compared to the
  amplitude of the driven motion due to the electric field component of
  the parametric drive. The presence of an electric field at the trapping position due
to the parametric drive causes a force on the atom which can be characterized and managed
in a fairly straightforward way.

%% Despite the usefulness of pulse shaping in this application, the time required 
%% to satisfy the adiabaticity criterion is set by the magnitude of the electric field
%% at the trapping position. Therefore, a limit on the mode-mode coupling strength is set
%% by the necessity to avoid unreasonably long pulse durations as the parametric drive
%% amplitude is increased.

The driven motion amplitude may be quantified by operating the
parametric drive in a continuous-wave mode. Then, the oscillating electric field at the 
ion position results in driven motion, analogous to the well-known
micro-motion~\cite{Wineland1998}. This driven motion causes the ion to
experience a frequency modulated laser field, redistributing
the laser power in frequency space and reducing the laser power at the
resonant frequency. This effect gives rise to sidebands around the
laser's carrier frequency at integer multiples of the driven motion
frequency.  For a continuous wave coupling field, the effect is
completely analogous to micro-motion leading to a reduced coupling
strength on the resonant optical transition which can be observed by a
decrease in the Rabi frequency $\Omega_c$. As in the case of
micro-motion, the optical transition can be driven by detuning the
laser by an integer multiple $n$ of the driving electric field
frequency. In that case, the transition strength is given
by~\cite{Berkeland1998}
\begin{align}
 \Omega_c \rightarrow |J_n ( k A)| \Omega_c \label{eqn:Bessel}
\end{align}
for a given oscillation amplitude $A$ along the laser propagation
direction $\vec{k}$ and $J_n$ being the n-th order Bessel function
of the first kind. We note that if the laser is not detuned, i.e. is on resonance
with the optical transition, the coupling strength is reduced by $J_0(k A)$.

\begin{figure}[htbp] 
	\begin{center}
	\includegraphics[width=1\columnwidth]{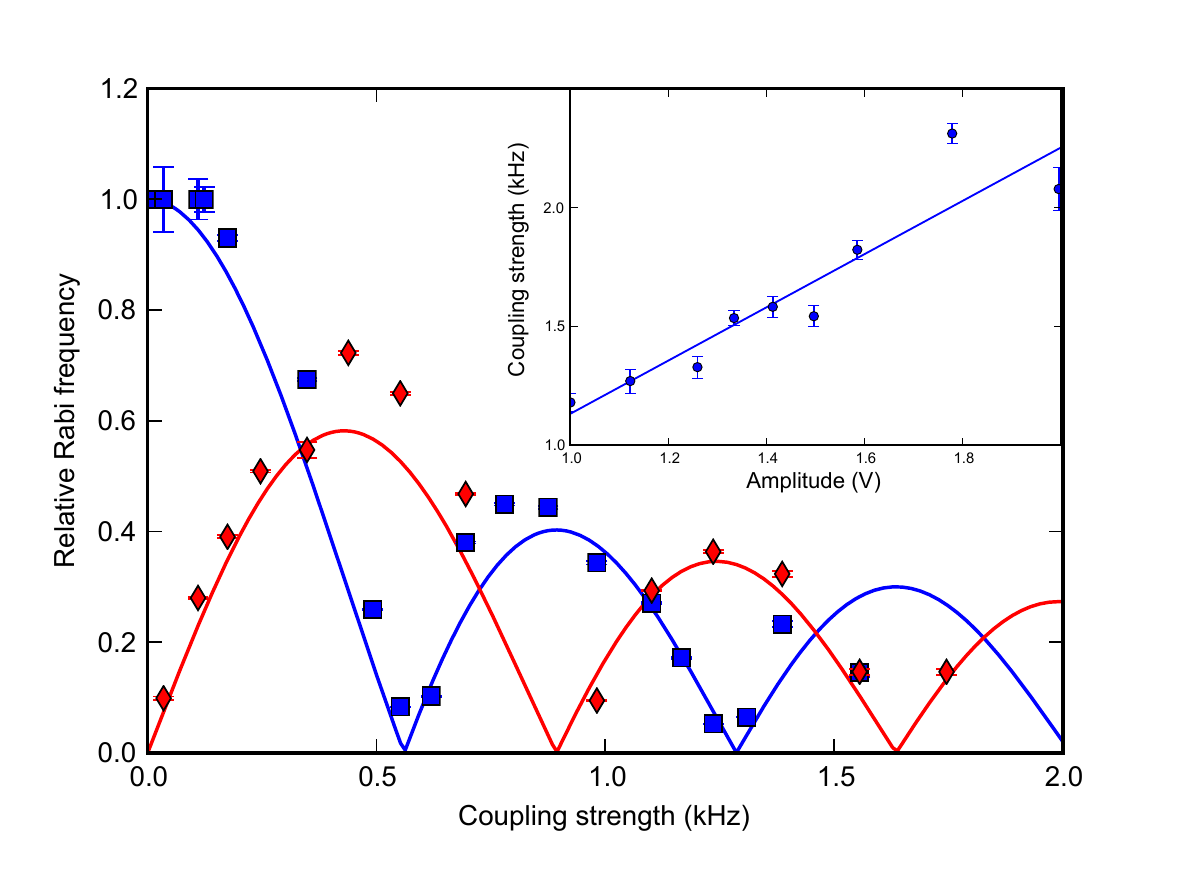}

        \caption{Relative Rabi frequency frequency on the carrier
          (blue squares) and driven motion sideband (red diamonds),
          compared to the unperturbed Rabi frequency on the
          unperturbed carrier, as a function of parametric coupling
          strength in the $xz$ coupling configuration. The solid
          lines represent fitted Bessel functions of the first
          kind. The inset illustrates the parametric coupling strength
          $g_{xz}$ as a function of drive voltage amplitude prior to
          the in-vacuum low pass filters. }
	\label{fig:bessel}
	\end{center}
\end{figure}

  To measure this effect, we apply a continuous drive (off resonant from all the 
  motional modes and first oder parametric resonances) onto the coupling
  electrodes and measure the frequency of Rabi oscillations on the 
  $| L, m_J\rangle = |S, -1/2 \rangle
  \rightarrow |D, -1/2\rangle$ transition. From this, we can extract
  the driven motion amplitude as a function of the parametric drive
  amplitude from Eq.~\ref{eqn:Bessel}. Fig.~[\ref{fig:bessel}].  shows
  the normalized Rabi frequencies on the carrier and the driven motion
  sideband as a function of the coupling strength for the xz coupling
  configuration where $\omega_p = 2\pi \times 1.7 \rm{\; MHz}$. This allows us to
  determine the ratio of driven motion amplitude to the coupling
  strength to be $A/g_{xz} = 497(8) \textrm{nm} /(2\pi \times 1 \textrm{kHz}$).

%The inset in figure \ref{fig:bessel} illustrates the
%coupling strength as a function of drive voltage, where the voltage is
%applied prior to the in-vacuum low-pass filters with a cut-off
%frequency of 300~kHz~\footnote{AVX X7R 47nF; part
%  no. W3H15C4738AT1F}. 

% Our measurements of the carrier transition strength for various
% drive amplitudes indicate that the driven motion along the laser axis should be about 60~nm per $2\pi \times1$~kHz of coupling. This discrepancy can be explained by an imperfection in the voltage ratio by which electrodes A and B (see Fig.~\ref{fig:trap}) are driven. Instead of the intended 1:4, a ratio of 1:4.7 is sufficient to explain the result.

In the xy configuration, the drive is applied to an electrode directly
beneath the ion so that most of the driven motion is in the direction
orthogonal to the laser and therefore does not significantly affect the
optical coupling strength.

%\begin{figure}[htbp] 
%	\begin{center}
%	\includegraphics[width=3in]{pulse_sequence.pdf}
%	\caption{Qualitative description of several experiments described in this work. Time is not to scale. Light grey sections indicate periods when the 729 nm laser is switched on at the ion. Black sections indicate that the parametric drive is switched on. In these experiments, the laser is always turned off when the parametric drive is operated. $\tau$ indicates a variable free evolution time between the state preparation and the Rabi excitation pulse, during which all lasers and the parametric drive are switched off. Doppler cooling and state readout are omitted. Details of the experiments are in the text.
%	}
%	\label{fig:pulse_sequence}
%	\end{center}
%\end{figure}

\subsection{Pulsed mode operation of the drive}
In the remainder of this work, we will investigate parametric coupling in the
pulsed mode. If the parametric drive is switched on and off rapidly, the micro-motion
analogy of Eq.~\ref{eqn:Bessel} no longer holds, and the unwanted
electric field can induce considerable off-resonant excitation in the oscillator
modes, disturbing the coupling dynamics. However, this electric field contribution to the total Hamiltonian
has no notable influence on the coupling dynamics if it is switched on
and off slowly enough, i.e. it is adiabatic. Here, the criterion for
adiabaticity is to avoid off-resonant excitation of the oscillator mode
itself. Experimentally, we shape the coupling field strength with a
Blackman windowing function, which has proven to effectively reduce
off-resonant excitation in a two-level
system~\cite{Harris1978,Riebe2006}. More precisely, the window for a
pulse with duration $T$ is described by
\[
B_T(t) = \frac{1-\alpha}{2} - \frac{1}{2} \cos\left(2\pi \frac{t}{T}\right) + 
\frac{\alpha}{2} \cos\left(4\pi \frac{t}{T}\right) 
\]
where $\alpha=0.16$. In order to facilitate the comparison to
rectangular pulses, the coupling duration of a Blackman shaped pulse
$B_T$ is defined as the duration of a rectangular pulse $T_{\rm{{rect}}}$
with the same pulse area so that $T = T_{\rm{rect}}/0.42$. Experimentally, using these pulses
for the xz configuration reduces the
  off-resonant excitation to less than 0.3 quanta for a reasonable
  coupling strength of several kHz.

\subsection{Population swapping}

\begin{figure*}[htbp] 
	\begin{center}
	\begin{tabular}{cc}
	\includegraphics[width=1 \textwidth]{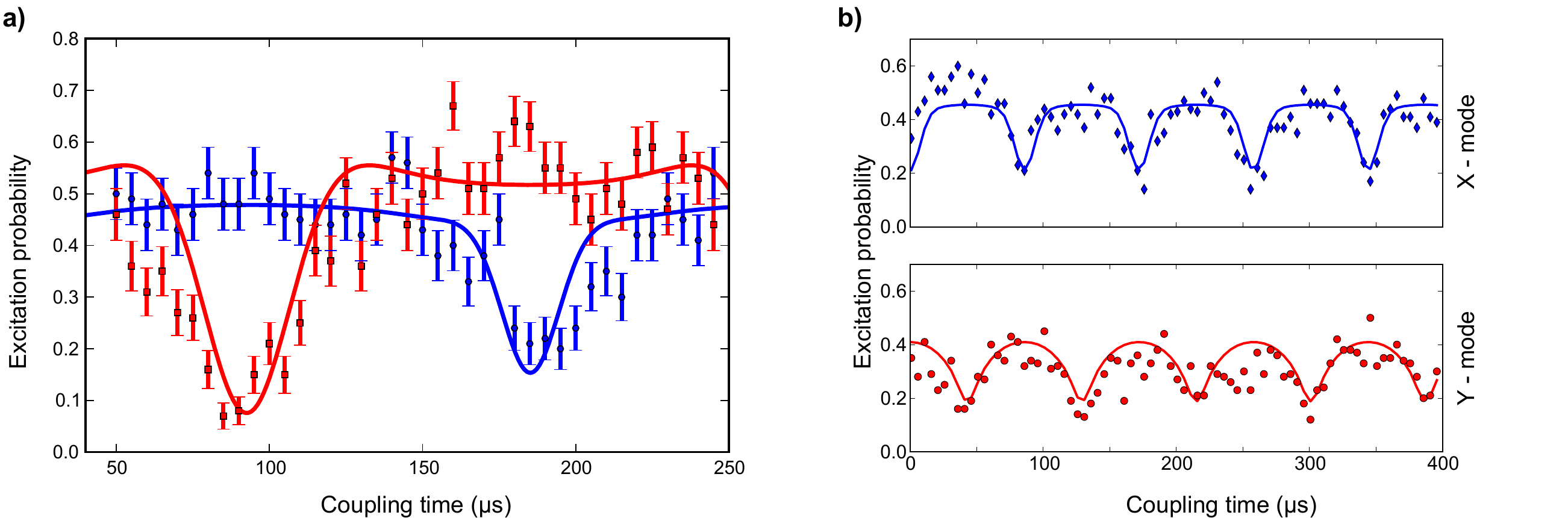} 
	\end{tabular}
        \caption{ (\textbf{a}) Time evolution of the coupling dynamics
          illustrated by the excitation of the red sideband of the Z
          (blue circles) and X (red rectangles) mode. Initially, the Z
          mode is cooled close to its ground state at a mean phonon
          number of $\bar{n}_z \approx 0.2$ while the X mode is left at the Doppler
          temperature of $\bar{n}_x \approx 6$ quanta. After a coupling time of
          around 90~$\mu$s, the population of the two modes is swapped
          and thus the X mode is close to its ground state. Solid
          lines correspond to a numerical solution of the model with
          no free parameters. The initial X mode population and coupling
          frequencies were determined from sideband spectroscopy.
          Note that the coupling time does not
          start at zero, because the Blackman shaped pulse is not
          adiabatic in this regime.  (\textbf{b}) Red sideband
          excitation of the X (blue) and Y (red) modes. The X mode is
          initially cooled to a mean phonon number of $\bar{n}_x
          \approx 0.3$. Solid lines indicate a fit to a model where
          the initial Y mode population and Rabi fequency are adjusted.
          The out of phase oscillations between the X
          and Y red sideband excitations show population
          oscillating between the two modes.  }
	\label{fig:swapping}
	\end{center}
\end{figure*}

 The first analysis in pulsed mode is to demonstrate that population can be exchanged
between two motional modes. This will serve as an experimental
definition of the exchange operation (SWAP), and form the cornerstone
for the cooling and analysis techniques presented later. To facilitate optical
analysis of both involved motional modes, we focus here on population
swapping in the xz configuration, but note that one can construct a
SWAP operation between any two modes and show as an example swapping
in the xy configuration.

To demonstrate population swapping, a single mode was sideband cooled
close to its ground state. For the xz configuration, the Z mode was
cooled, and for the xy configuration, the X mode was cooled, followed by a 
mode coupling pulse, applied for a variable time. The motional
state after the coupling was probed on the red sideband of either mode
on the $|S, -1/2 \rangle \rightarrow |D, -1/2\rangle$ transition.  As
the mean phonon number in a given mode drops significantly below one,
the excitation probability is
suppressed\cite{Wineland1998,leibfried2003}. The dynamics of the
coupled systems are illustrated in Fig. [\ref{fig:swapping}] where the
periodic oscillations of the excitation probability represent a
hallmark feature of the coupling. It furthermore allows us to define a
SWAP operation where the state of the two modes are completely
exchanged at around 90~$\mu$s for xz coupling, and 50~$\mu$s for xy
coupling. For the xz coupling configuration, a Blackman shaped pulse
needs to be used whereas for the xy configuration a square pulse is
sufficient to suppress off-resonant excitation.

%%%%%%%%%%%%%%%%%%%%%%%%%%%%%%%%%%%%%%%%%%%%%%%%%%%%%%%%%%%%%%%%
%%%%%%%%%%%% COOLING WITHOUT DIRECT ACCESS %%%%%%%%%%%%%%%%%%%%%
%%%%%%%%%%%%%%%%%%%%%%%%%%%%%%%%%%%%%%%%%%%%%%%%%%%%%%%%%%%%%%%

\section{Cooling without direct access} 

\subsection{Application: Ground state cooling}

The parametric coupling technique we describe here provides a useful
addition to the mostly optical toolbox in ion trapping physics. In
particular, it enables cooling motional modes without accessing them
by laser beam.

The basic principle involves performing laser cooling on a single,
laser-accessible mode (the {\em primary mode}), followed by population
swapping to transfer energy from a secondary, uncooled mode, into the
primary mode. One implementation of this technique is to perform
several cooling cycles on the primary mode, and to insert a SWAP
operation between each cycle. With the parametric interaction operated
in this way, the primary mode provides a cold reservoir for the
secondary mode. After each cycle of sideband cooling, the populations
of the primary and secondary modes are swapped, eventually resulting
in a state where both the primary and secondary modes are prepared
close to their ground state. We call this method of cooling {\em interleaved cooling}, 
allowing us to prepare both modes close to
their ground states. Interleaved cooling is particularly elegant for
ground state preparation because it is insensitive to errors in the
SWAP operation. Even a somewhat imperfect SWAP operation will transfer a
large fraction of the population between the two modes mode, where the
population in the primary mode is then removed by optical
cooling. Repetition of this process several times leads to significant
reduction in the secondary mode population.

  Interleaved cooling can prepare both modes close to
  their ground states when the heating processes on both modes are
  slower than the cooling rate on the primary mode. However, even if
  this condition is not satisfied,  it is
  still possible to prepare the secondary mode close to its ground
  state. Here, we take advantage of the SWAP operations
  that can be performed much faster than the typical cooling processes on
  optical transitions, as one is not limited by the relatively weak
  coupling of the light to the ion's motion. Therefore, one can cool
  the primary mode and perform a \emph{single SWAP} operation
  subsequently, resulting in a cool secondary but a hot primary
  mode. This method is applicable as long as the SWAP operation can be
  performed faster than any heating process on the secondary mode.

%  If
% this condition is not satisfied, the secondary mode may acquire more
% than one thermal quantum during cooling on the primary mode. This will
% make it impossible to simultaneously cool both the primary and
% secondary modes to the ground state. In this regime, it is still
% possible to prepare the secondary mode into the ground state. To do
% this, one employs a {\em single SWAP} operation after sideband cooling
% the primary mode. This operation then leaves the secondary mode with a
% small thermal occupation. In this way, modes which heating rate is
% higher than the sideband cooling rate can be easily prepared in the
% ground state as long as the SWAP operation is short compared to the
% heating processes.
% }

We have tested both of these cooling techniques, using the Z mode as
the primary and the X mode as the secondary. In the case of
interleaved cooling, we performed eight cycles of sideband cooling
with a SWAP operation between each, wheras for the single SWAP cooling, the Z mode
was cooled for 8~ms followed by a single SWAP. We tabulate the results
detailed in Table~I, showing that both simultaneous ground state
cooling by interleaved cooling, as well as single swap cooling are
effective techniques for cooling.

\begin{table}[t]
\begin{tabular} {l | | c | c}
	\textbf{Method} & $\bar{n}_z$ & $\bar{n}_x$ \\
        \hline
	Interleaved & 0.13(2) & 0.31(5) \\
	Single SWAP & 7(5) & .7(2)
\end{tabular}
\label{table:cooling}
\caption{Cooling results for both interleaved and single SWAP cooling methods.
The interleaved method is capable of preparing two motional modes of the ion
close to the ground state
, provided that the heating rate in both modes
is sufficiently low (see text). That the single SWAP method is somewhat less effective
than the interleaved method for ground state preparation reflects the method's higher
sensitivity to errors in the parametric resonance frequency and
mode swapping time as compared to the interleaved scheme.
}
\end{table}

\subsection{Application: Heating rate of an inaccessible mode}

A second way to use the parametric interaction is to probe the thermal occupation of an inaccessible mode. We show that a heating rate can be measured in a mode nearly orthogonal to the laser propagation direction.
%
% For thermally occupied modes with an appreciable ground state population (that is, modes with an average excitation of up to a few quanta), the temperature can be easily extracted by comparing optical excitation on the red and blue secular sidebands. When a mode is in the ground state, the red secular sideband vanishes entirely, and therefore optical spectroscopy of modes with small thermal excitation will feature a red sideband significantly smaller than the blue sideband. 
%

The heating rate on a single mode can be accurately determined by a process of cooling the mode to an average excitation much smaller than one vibrational quantum, and then probing the red and blue sideband excitation as a function of a variable waiting time after cooling\cite{leibfried2003}. However, this process relies on the ability to prepare the motional state to small mean phonon numbers, as well as optical access to the secular sidebands to read out the mode occupation.

In our experiment, the Y mode lies nearly perpendicular to the plane of the trap, such that the projection of the laser onto this mode is about $9^{\circ}$--too small to use sideband cooling directly on the mode. Thus, in order to prepare the mode to small thermal occupation, we performed sideband cooling on the x-mode and then a single SWAP operation to initialize the y-mode to a mean thermal occupation of less than a single quantum. To determine the heating rate, we analyze the mode temperature as discussed above after a variable waiting time.

Due to the fact that the laser is not completely orthogonal to the
y-mode, the mode temperature can be analyzed directly on the secular
sidebands corresponding to the Y secular sideband. However, this
requires comparably long (exceeding 500 $\mu$s) optical excitation
times, during which instabilities in the mode frequency cause
systematic errors. Furthermore, the excitation time is not short
compared to the heating processes, adding another systematic error.
Therefore, a much cleaner approach is to again apply a SWAP operation
between the X and Y modes to exchange their population.

\begin{figure}[htbp]
	\begin{center}
	\includegraphics[width=1 \columnwidth]{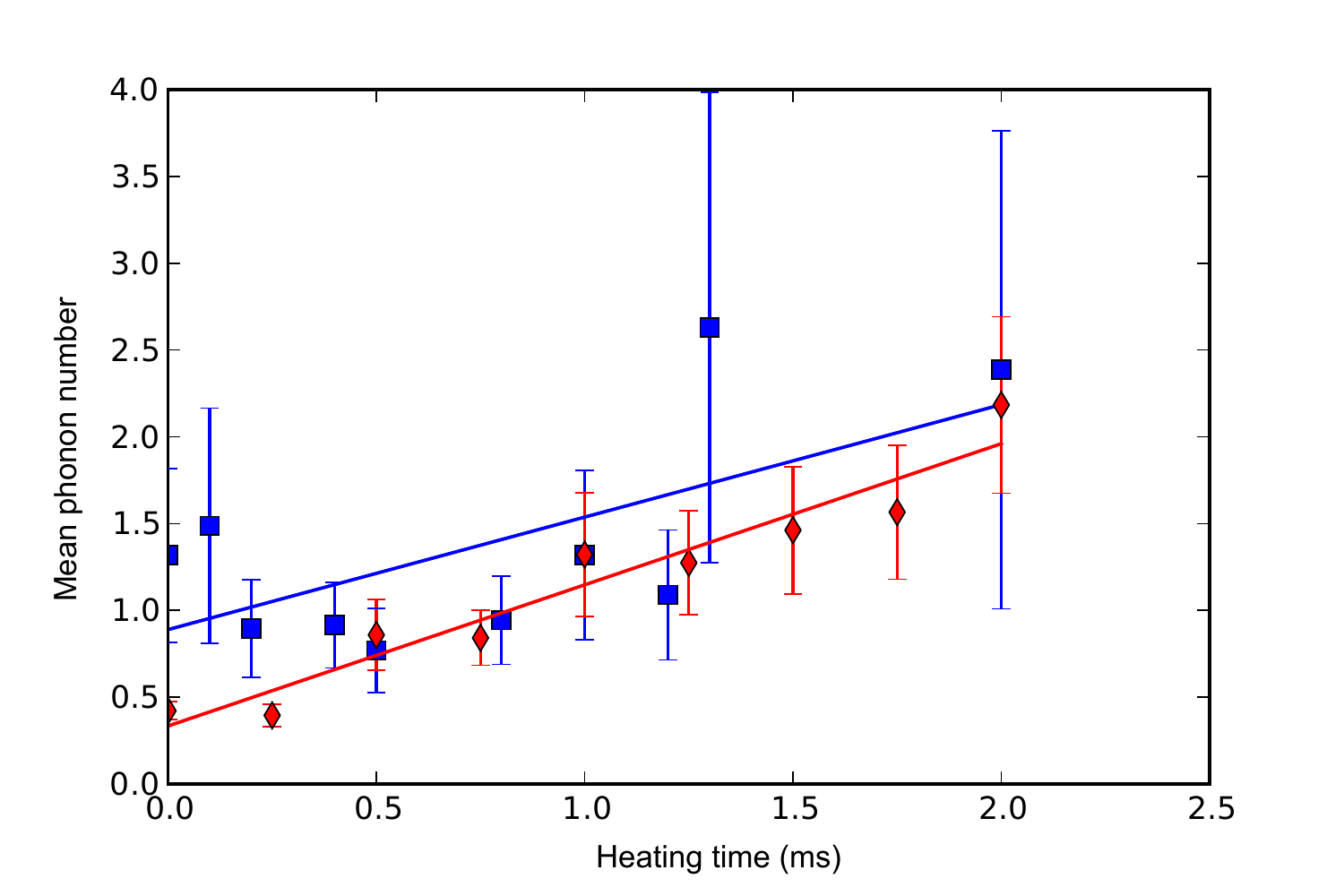}

        \caption{Mean phonon number as a function of heating time on
          the Y mode. The Y mode is prepared close to its ground state by
          cooling the X mode and swapping the motional
          states. Analysis of the motional state is either performed
          directly on the Y mode (blue squares) or by a second
          coupling operation and subsequent analysis on the X mode
          (red diamonds). The red line corresponds to a heating rate of
         810(80) quanta/s, while the blue line corresponds to a heating rate
        of 650(270) quanta/s.}
	\label{fig:heating_rate}
	\end{center}
\end{figure}

In order to verify that the second SWAP operation works as expected,
we performed a heating rate experiment using both of the above
techniques, with the heating time varied between 0 and 2 ms. The
results of these experiments are shown in
Fig. [\ref{fig:heating_rate}]. The measurement using the direct
analysis of the sideband infers a heating rate of 650(270) quanta/s
whereas the measurement employing two SWAP operations yields 810(80)
quanta/s.  The smaller uncertainty from temperature of the Y mode via
the X mode reflects the fact that the Rabi frequency on the X red
sideband is considerably faster than that of the Y mode. This renders the temperature
measurement much less sensitive to instabilities in the radial
motional frequencies due to the reduced Fourier bandwidth of the
applied pulses.

\section{Conclusions}

We have shown that inducing mode-mode coupling by means of a parametric drive can be readily applied to the case of trapped ions in surface electrode Paul traps. The presented method provides control of motional modes which lack direct optical access, reducing experimental design constraints. Both in surface science studies as well as experiments aimed at coupling
the motion of single ions to solid state systems~\cite{Daniilidis2013b,Daniilidis2013,Daniilidis2009a,Hite2012}, optical access to the modes of interest might be seriously compromised. Thus, we expect this
parametric coupling technique to be useful in overcoming the related challenges of these experiments
by allowing the experimenter to control and measure the temperature of optically inaccessible
modes.

We also used the mode-mode coupling scheme to show ground state cooling and a subsequent heating rate measurement on a mode that was not addressed optically. With heating rates sufficiently slow, the interleaved cooling scheme can be used to simultaneously prepare two modes close to their quantum mechanical ground state. Simultaneous cooling of several modes has been shown for instance in electromagnetically induced transparency (EIT) cooling, but in that case the modes must be sufficiently close in frequency for cooling to be effective~\cite{Roos2000a}. The interleaved cooling technique, however, can also be applied if the modes have very different frequencies.

It may also be possible to use this method to generate non-classical
motional states. Indeed, as we noted in
Eq.~\ref{eqn:hamplifier_rwa}, the system can be operated as a
parametric amplifier, yielding two-mode
squeezing~\cite{Gerry2004Introductory}. Additionally, it may be
possible to generate two-mode entanglement by ground state cooling two
modes, performing coherent rotation on a single mode, and then
applying the driving pulse for half of the swap time.

\section{Acknowledgements}

This work has been supported by AFOSR through the ARO grant FA9550-11-1-0318.

%Merlin.mbs v4.21 2009-07-09.
%

%\bibliography{paper,schroedinger}
%\bibliography{schroedinger}
%\bibliographystyle{my_apsrev4-1}

\end{document}